# Bubble Cloud Characteristics and Ablation Efficiency in Dual-Frequency Intrinsic Threshold Histotripsy

Connor Edsall[1], Laura Huynh[2], Tim Hall[3], Eli Vlaisavljevich[1,4]

1 - Department of Biomedical Engineering and Mechanics, Virginia Polytechnic Institute and State University, 325 Stanger St., Blacksburg, VA, 24061
2 - Department of Materials Science and Engineering, Virginia Polytechnic Institute and State University, 445 Old Turner St., Blacksburg, VA 24061
3- Biomedical Engineering, University of Michigan, Carl A. Gerstacker Building, 2200 Bonisteel Blvd, Ann Arbor, MI 48109-2133, USA
4 - ICTAS Center for Engineered Health, Virginia Polytechnic Institute and State University, 325 Stanger St., Blacksburg, VA, 24061

**Corresponding Author: Eli Vlaisavljevich**
Virginia Polytechnic Institute and State University
Department of Biomedical Engineering and Mechanics
325 Stanger Street, Blacksburg, VA 24061
Phone: 540-231-2136
E-mail: eliv@vt.edu

# Abstract

Histotripsy is a non-thermal focused ultrasound ablation method that destroys tissue through the generation and activity of acoustic cavitation bubble clouds. Intrinsic threshold histotripsy uses single-cycle pulses to generate bubble clouds when the dominant negative pressure phase exceeds an intrinsic threshold of ~25-30 MPa. The ablation efficiency is dependent upon the size and density of bubbles within the bubble cloud. This work investigates the effects of dual-frequency pulsing schemes on the bubble cloud behavior and ablation efficiency in intrinsic threshold histotripsy. A modular 500 kHz:3 MHz histotripsy transducer treated agarose phantoms using dual-frequency histotripsy pulses with a 1:1 pressure ratio from 500 kHz and 3 MHz frequency elements and varying arrival times for the 3 MHz pulse relative to the arrival of the 500 kHz pulse (-100 ns, 0 ns, and +100 ns). High-speed optical imaging captured cavitation effects to characterize bubble cloud and individual bubble dynamics. The effects of dual-frequency pulsing on lesion formation and ablation efficiency were also investigated in red blood cell (RBC) phantoms. Results showed that the single bubble and bubble cloud size for dual-frequency cases were intermediate to published results for the component single frequencies of 500 kHz and 3 MHz. Additionally, bubble cloud size and dynamics were shown to be altered by the arrival time of the 3 MHz pulse with respect to the 500 kHz pulse, with more uniform cloud expansion and collapse observed for early (-100 ns) arrival. Finally, RBC phantom experiments showed that dual-frequency exposures were capable of generating precise lesions with smaller areas and higher ablation efficiencies than previously published results for 500 kHz or 3 MHz. Overall, results demonstrate dual-frequency histotripsy's ability to modulate bubble cloud size and dynamics can be leveraged to produce precise lesions at higher ablation efficiencies than previously observed for single-frequency pulsing.

# Introduction

Histotripsy is a non-invasive and non-thermal focused ultrasound ablation method currently being developed for the treatment of cancer and other clinical applications (Bader *et al.*, 2019; Xu *et al.*, 2021; Vidal-Jove *et al.*, 2022). Histotripsy mechanically destroys tissue through the precise control of acoustic cavitation (Vlaisavljevich *et al.*, 2013b; Vlaisavljevich *et al.*, 2016; Bader *et al.*, 2019). High-pressure (> 10 MPa), short-duration (< 20 µs) pulses applied at very low duty cycles (< 1 %) produce histotripsy "bubble clouds" (Parsons *et al.*, 2006; Roberts *et al.*, 2006; Xu *et al.*, 2004). The high stress and strain induced in the target tissue by the rapid expansion and collapse of the individual bubbles comprising the bubble cloud result in the destruction of cells and the surrounding extracellular matrix (Hall *et al.*, 2009; Vlaisavljevich *et al.*, 2013b; Vlaisavljevich *et al.*, 2016). Complete ablation of the focal volume following repeated pulsing is characterized by well-defined lesions closely matching the size and location of the bubble cloud (Roberts *et al.*, 2014; Vlaisavljevich *et al.*, 2017; Zhang *et al.*, 2015), with the necessary number of pulses dependent on the tissue's mechanical properties and bubble cloud characteristics (Vlaisavljevich *et al.*, 2015c; Mancia *et al.*, 2017; Vlaisavljevich *et al.*, 2017; Edsall *et al.*, 2021a).

Intrinsic threshold histotripsy is one of three currently established histotripsy methods, along with shock scattering histotripsy and boiling histotripsy (Bader *et al.*, 2019). Intrinsic threshold histotripsy distinctly forms acoustic cavitation directly from the dominant, high negative-pressure phase of a single pulse (≤ 2 cycles) with generation dependent upon the duration and amplitude of the negative pressure as well as the target medium's material properties (Maxwell *et al.*, 2013; Vlaisavljevich *et al.*, 2015b). The intrinsic threshold for cavitation generation has consistently been shown to be ~25-30 MPa for water-based soft tissues for histotripsy pulse frequencies ranging from 345 kHz to 3 MHz (Vlaisavljevich *et al.*, 2015b; Maxwell *et al.*, 2013; Edsall *et al.*, 2021a). Acoustic cavitation near the threshold presents as individual bubbles with the characteristic histotripsy bubble cloud forming with increasing pressure (Vlaisavljevich *et al.*, 2015b; Vlaisavljevich *et al.*, 2017). The bubble cloud's dimensions closely match the volume predicted by the focal region above the intrinsic threshold (Lin *et al.*, 2014b).

Pulsing parameters have experimentally been shown to affect bubble cloud characteristics in intrinsic threshold histotripsy. For instance, prior work has shown that decreasing transducer f-number increases the bubble density within the bubble cloud, with a corresponding increase in ablation efficiency (Vlaisavljevich *et al.*, 2017). Studies have also shown that decreasing pulse frequency leads to increased individual bubble expansion and bubble cloud size due to the longer duration of the suprathreshold negative pressure ($p$-) (Vlaisavljevich *et al.*, 2015b; Mancia *et al.*, 2017). As a result of these changes in bubble cloud characteristics, lower frequency correlates with an increased ablation efficiency, which was hypothesized to be due to increased bubble expansion inducing higher strain on the targeted tissue (Edsall *et al.*, 2021a). However, lower-frequency pulsing also produces clouds with reduced bubble density (Edsall *et al.*, 2021a), suggesting that further increases in ablation efficiency could be achieved using methods capable of simultaneously increasing bubble density as well as individual bubble expansion.

Dual-frequency histotripsy pulsing strategies have previously been developed to modulate bubble cloud size and individual bubble expansion (Lin *et al.*, 2014a; Vlaisavljevich *et al.*, 2015a). These prior studies have shown that the size of the bubble cloud and individual bubble size can be modulated using dual-frequency pulsing strategies by changing the percentage of pressure coming from the respective frequencies (Vlaisavljevich *et al.*, 2015c; Lin *et al.*, 2015; Lin *et al.*, 2014a). These prior works have also shown that the size of a single histotripsy bubble and the corresponding region of tissue ablation were not affected by the arrival times of the respective frequencies as long as the peak negative pressures remained above the intrinsic threshold, which was shown to occur for cases in which the high frequency 3 MHz pulse arrived within 0.15 μs relative to the low frequency 500 kHz pulse (Lin *et al.*, 2014a). However, these prior studies did not characterize the bubble cloud characteristics and behavior in dual-frequency histotripsy, nor did they assess the effects of the pulse arrival time on the bubble cloud density and ablation efficiency for higher peak negative pressures ($p$-) that are commonly used in histotripsy to generate bubble clouds instead of individual bubbles, which is critical to understanding and optimizing dual-frequency histotripsy pulsing methods for specific clinical applications.

Using the same modular transducer and similar methods as our prior investigation of single-frequency (Edsall *et al.*, 2021a), this study investigates the effects of dual-frequency histotripsy pulsing on the characteristics, behavior, and ablation efficiency of histotripsy bubble clouds generated inside agarose tissue phantoms. This study also compares the effects of 3 MHz pulse arrival modulation relative to the 500 kHz pulse arrival on the bubble cloud characteristics, behavior, and ablation efficiency. Cavitation activity was captured using two modes of high-speed optical imaging to compare the single-bubble and bubble cloud dimensions, behavior, and the resulting bubble cloud ablation efficiency for dual-frequency (500 kHz:3 MHz) cases.

The primary hypothesis of this study is that dual-frequency pulsing can be used to modulate the bubble cloud characteristics, behavior, and ablation efficiency in intrinsic threshold histotripsy distinctly from its component frequencies. We expect the results to show that the bubble cloud size and individual bubble size will be intermediate to previously published values for the component 500 kHz and 3 MHz single frequencies across all 3 MHz arrival times. The result of this will be that dual-frequency pulsing will produce intermediately sized lesions compared to previously published results for 500 kHz and 3 MHz single frequencies. We further hypothesize that the arrival time of the high-frequency (3 MHz) pulses relative to the arrival of the low-frequency (500 kHz) pulses will impart additional degrees of frequency-specific characteristics on the nucleation and dynamics of the resulting acoustic cavitation. Based on our prior investigation of the effects of single frequencies, we hypothesized that bubble cloud size and resulting lesion dimensions would be intermediate to the single frequency cases, with a slightly smaller bubble cloud formed with an earlier arrival of the 3 MHz waveform corresponding to a smaller focal area above the threshold. We further hypothesized that dual-frequency histotripsy would result in an increased ablation efficiency (evaluated as the number of pulses required to remove the area of the bubble cloud) as these bubble clouds should be denser bubble clouds (due to higher frequency nucleation) comprised of bubbles with larger expansion (due to the elongated tensile phase of lower frequency component). Overall, the findings of this study will increase our understanding of how dual-frequency pulsing techniques can be used to modulate bubble cloud characteristics and resulting ablative effects in histotripsy.

## Methods

*Ultrasound Pulse Generation*

To directly compare the results of this study with our prior work investigating the effects of frequency on histotripsy cavitation and behavior (Edsall *et al.*, 2021a), all experiments used the same custom-designed, modular histotripsy transducer with a geometric focus of 75 mm, an aperture size of 120.5 mm, and a f-number of 0.62 (**Fig. 1B**). Sixteen 500 kHz and sixteen 3 MHz elements (alternating every other element) populated the transducer scaffold which was comprised of three concentric rings of six, twelve, and fourteen ports (**Fig. 1B**). The transducer was positioned horizontally in a tank of DI water which was degassed to an oxygen concentration <28 % O2. A custom, high-voltage amplifier populated with interchangeable amplifier boards matched to the frequency (500 kHz or 3 MHz) of the respective elements produced short therapy pulses of <2 cycles. A field-programmable gate array (FPGA) board (Altera DE0-Nano Terasic Technology, Dover, DE, USA), programmed by a custom MATLAB (The MathWorks, Natick, MA, USA) code, synchronized the pulse arrivals by controlling the phase-delay and the charge-times of each individual element for each frequency. For each of these experiments, a MATLAB code coordinated the motion of a three-axis positioning system to orient agarose tissue phantoms or hydrophones with the pulsing of the transducer for the respective experiments (**Fig. 1A**).

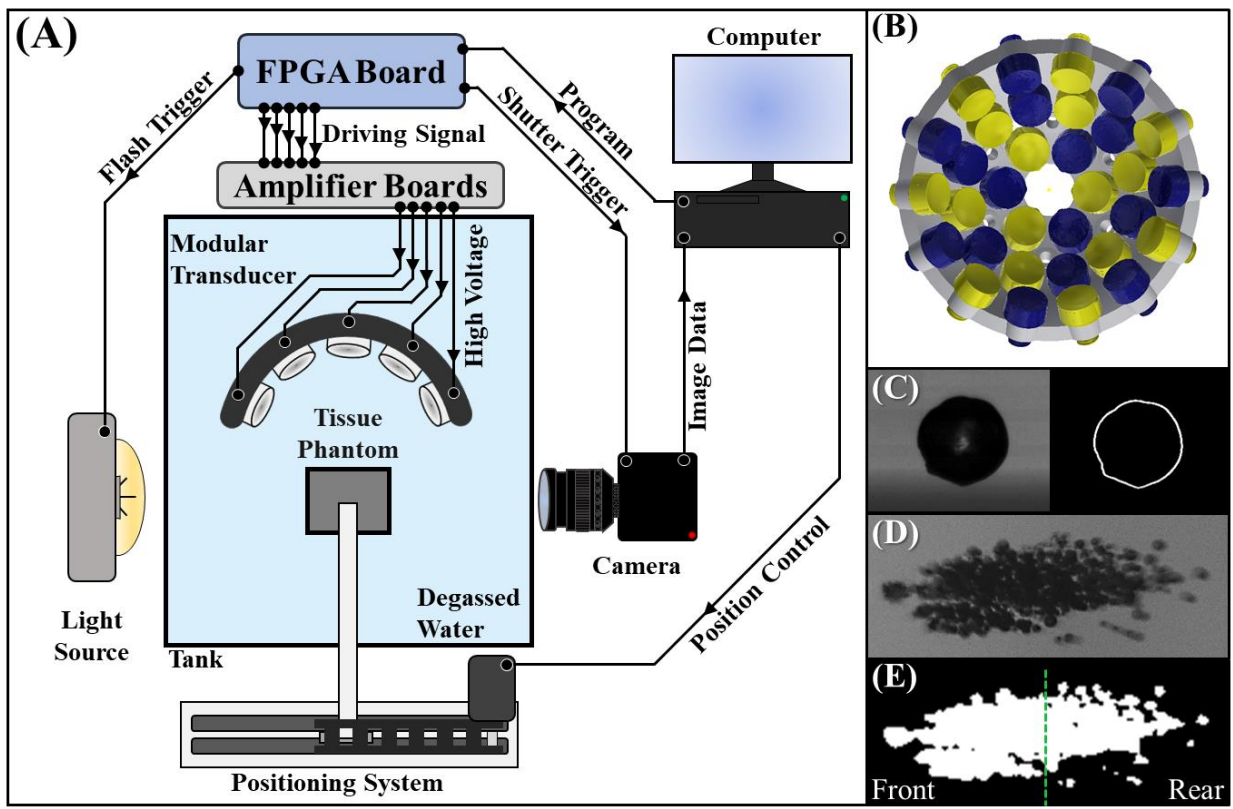

*Figure 1 Dual-Frequency Histotripsy Experimental Schematic.* **(A)** Optical imaging using a FLIR camera and high-speed strobe captured cavitation nucleation and red blood cell (RBC) ablation inside agarose tissue phantoms exposed to histotripsy pulses applied by the 32-element modular transducer **(B)** with 500 kHz (blue) and 3 MHz (yellow) arranged in an alternating pattern. High-speed imaging using a Phantom high-speed camera captured single bubble and bubble cloud dynamics. **(C)** Single bubbles were captured in greyscale videos and traced in binary to measure the bubble expansion and collapse. **(D)** For bubble cloud behavior, each bubble cloud was captured in greyscale, converted to binary, and the centroid of the cloud located. **(E)** The cloud area of growth and collapse was then measured in total as well as before (front) and after (rear) the centroid of the cloud in the axial dimension (green dotted line**)**.

*Hydrophone Focal Pressure Calibration*

A fiberoptic probe hydrophone (FOPH), cross-calibrated with a high-sensitivity reference rod hydrophone (HNR-0500, Onda Corporation, Sunnyvale, CA, USA), measured the dual-frequency pulsing, which consisted of a 1:1 pressure amplitude of 500 kHz:3 MHz elements. The 500 kHz and 3 MHz peaks were first aligned such that the *p-* of each frequency arrived at the focal point concurrently (0 ns offset) with the *p-* resulting from the linear addition of the *p-* of the individual 500 kHz and 3 MHz waveforms (**Fig. 2**). The FOPH was then also used to determine the necessary phase delay for each element to have the 3 MHz waveform arrive ± 100 ns relative to the peak of the 500 kHz (**Fig. 2**). The FOPH directly measured the waveforms in degassed water up to peak negative pressures (*p-*) of $p- = 18$ MPa for each arrival time. Higher focal pressures, up to ~36 MPa, were estimated by summing the outputs from the transducer sub-apertures of sixteen elements (eight 500 kHz and eight 3 MHz) to avoid cavitation damage to the hydrophone. An oscilloscope (TBS2000 series, Tektronix, Beaverton, OR, USA) collected waveforms at a sample rate of 500 MS/s. The waveforms were averaged over 512 pulses to minimize signal noise and recorded to MATLAB.

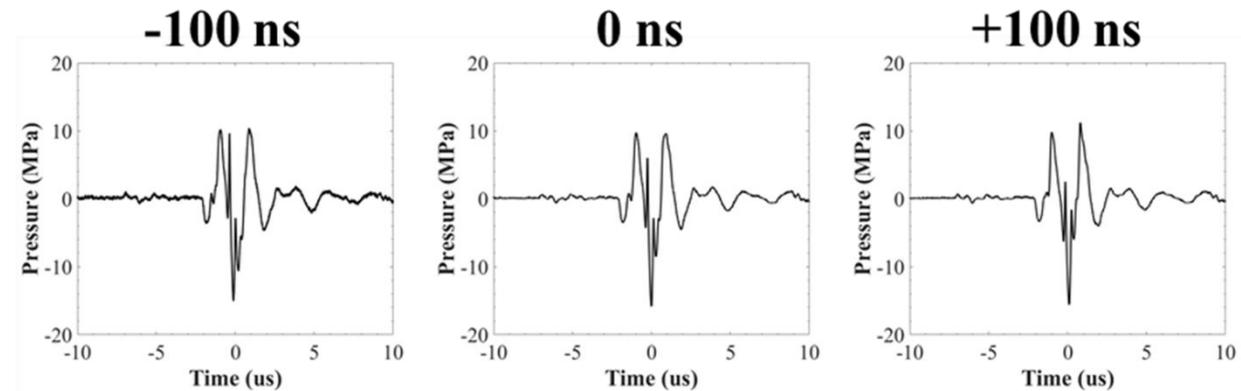

*Figure 2 Example Dual-Frequency Waveforms.* 3 MHz pulse arrivals were set to -100 ns, 0 ns, and +100 ns relative to the arrival of the 500 kHz pulse. Example acoustic waveforms showing the combined 500 kHz:3 MHz dual-frequency waveform at the focus for each 3 MHz arrival time.

*Agarose Gel Phantom Preparation*

Agarose preparation followed previously published methods (Edsall *et al.*, 2021a; Edsall *et al.*, 2021b; Vlaisavljevich *et al.*, 2014). 1 % (w/v) agarose tissue phantoms with a Young's modulus of ~21.7kPa (Vlaisavljevich *et al.*, 2015c) were used to replicate soft human tissue's mechanical and viscoelastic properties (Duck, 1990; Yamada, 1973). To make the tissue phantom, the agarose gel was prepared by mixing 0.5 % agarose powder (Type VII-A, Sigma Aldrich Corporation, St. Louis, MO, USA) with 99.5 % of 0.45 % saline at room temperature. Saline was prepared from a stock 4.5 % saline and DI water degassed using a Portable Water Degasser (DS-50, FUS Instruments, Toronto, ON, CAN). The mixture was heated in a microwave until boiling and then stirred to fully dissolve the agarose. Next, the sample was repeatedly heated to boiling and stirred to produce flash boiling to release dissolved gas from the mixture until the solution was fully degassed and 50 % of the volume remained, leaving a degassed 1 % (w/v) agarose gel mixture. The mixture was placed under a partial vacuum (~33.62 kPa, absolute) for twenty minutes to remove the remaining gas and minimize regassing as the agarose solution cooled. The pressure was decreased to ~16.75 kPa for an additional five minutes to force any remaining gas from solution. Once the temperature of the agarose dropped to 40 ºC, a serological pipette was used to slowly inject 120 mL of the gel down the wall of a rectangular silicone mold into a custom-designed polylactide (PLA) phantom holder frame. The silicone mold containing the gel was then stored in a refrigerator for one hour to solidify. Each experiment was performed within two hours of the gel creation to ensure a consistent agarose concentration and the degassed state of the gel during testing.

*Optical Imaging for Characterization of Bubble Cloud Size*

Optical imaging using a high-speed camera (FLIR Blackfly S monochrome, BFS-U3-32S4M-C 3.2 MP, 118 FPS, Sony IMX252, Mono, FLIR Integrated Imaging Solutions, Richmond, BC, Canada) and a 100 mm F2.8 Macro lens (Tokina AT-X Pro, Kenko Tokina Co., LTD, Tokyo, Japan) monitored cavitation events and characterized the bubble clouds. Captured images using this combination had a resolution of 3.25 μm per pixel. A custom high-speed LED strobe light backlit each sample with the exposure duration kept as low as possible (1 μs) to minimize the motion blur of the expanding bubbles. Timing accuracy of the pulse delivery, camera and strobe timings, and all delays were accurate to the 10 ns clock speed of the FPGA. The FPGA in the amplifier box triggered the strobe and camera such that all exposures were centered at a 3 μs delay after the arrival of the peak negative pulse. This delay was optimized to detect clearly visible bubbles while minimizing bubble overlap within the bubble cloud. A 1 Hz pulse repetition frequency (PRF) was used for bubble cloud characterization experiments to limit the memory effect on subsequent pulses (Maxwell *et al.*, 2013). Images captured very early in the life of the bubble cloud at a low PRF minimized memory effect between pulses. The optical images for each sample were analyzed using a custom MATLAB script that identified bubble clouds for dimensional measurement by converting the captured grayscale image into and negative binary image based on an intensity threshold determined by the difference from the background intensity following previously published methods (Maxwell *et al.*, 2013).

The effect of the 3 MHz arrival in dual-frequency histotripsy on bubble cloud size was compared by capturing bubble clouds for each condition (-100, 0, +100 ns). For each dual-frequency condition, 100 histotripsy pulses were applied to agarose phantoms for each frequency and pressure level combination. The bubble cloud length was detected as the distance from the closest edge of the nearest bubble to the transducer to the furthest edge of the furthest bubble from the transducer. Similarly, the bubble cloud height was defined as the distance from the topmost edge of the bubble highest in the positive elevational direction to the bottom edge of the lowest bubble in the negative elevational direction. These dimensions were then used to calculate the mean cloud area using $A = \left(\frac{Axial\ Dimension}{2}\right)\left(\frac{Elevational\ Dimension}{2}\right)\pi$.

*Single Bubble and Bubble Cloud Dynamics*

A high-speed camera (Phantom High-Speed TMX-6410, Mono, 18.5 µpixel, Vision Research Inc., Wayne, NJ, USA) coupled to an 85 mm f/2.8 1-5x super-macro lens (Creator, Mitakon Zhongyi, Liaoning Province, CHN) monitored the growth and collapse behavior of single bubbles and bubble clouds in 1 % agarose phantoms. The phantoms were backlit by steady LED-array light source (LT-V9-15, GSVITEC, Bad Soden-Salmünster, DEU). The Phantom high-speed camera was set to image at 200,000 frames per second capturing 1 frame every 5 µs from -1 µs before until 274 µs after *p-* arrival. The maximum resolution at this speed was 512x256 pixels giving a per pixel resolution of ~5 µm.

For single-bubble dynamics, 50 pulses were applied to the phantom at 1Hz PRF at pressures just above their respective threshold each 3 MHz arrival time (-100, 0, +100 ns). These ranges were selected for analysis as they fell within a narrow range of delays (-150 ns to 150 ns) previously reported to have minimal effect on cavitation threshold and respective lesion size from single bubbles (Lin *et al.*, 2014a). The high-speed camera collected and saved 56 frames per pulse.

These frames were converted to an AVI file tracking the single-bubble growth and collapse. A custom MATLAB script traced the boundary of the bubble through each frame of the AVI file (**Fig. 1C**), marked the diameter in the axial and elevation directions, and reported the mean bubble radius for the bubble that would approximate a circular area at each 5 µs interval for each generated bubble. These dimensions were averaged for each 5 µs time point across for the first 25 single bubble (excluding instances of 0 bubbles and >1 bubble) producing the mean radius and standard deviation. The mean maximum radius, $R_{max}$, and the mean time, $t_{max}$, to reach that maximum bubble size were recorded. Mean collapse time the bubble before rebounding was also recorded ($t_{c1}$). Lastly, the final time point of collapse was recorded, $t_{c2}$, was recorded after a small rebound of each bubble was observed.

For bubble cloud dynamics, 30 pulses were applied to the phantom at 1 Hz PRF at 36 MPa for each 3 MHz arrival time (-100, 0, +100 ns). The high-speed camera collected and saved 56 frames per pulse. These frames were converted to greyscale JPEG images tracking the growth and collapse of the bubble clouds generated by each pulse (**Fig. 1D**). A custom MATLAB program created binary images of the greyscale cloud using thresholding to identify bubbles using prior methods (Edsall *et al.*, 2021a) similar to those listed in the characterization of bubble cloud size in section *3.2.4*. The centroid of the cloud captured in frame 2 (4 µs) was located by MATLAB and designated as the center point to divide all subsequent clouds into front and back regions (**Fig. 1E**). The code then recorded the total, front, and rear areas of each cloud for each 5 µs time point through growth and collapse. The area was averaged for each 5 µs time point across for all 30 bubble clouds producing the mean total, front, and rear areas, and the respective standard deviations for each area at each time point.

*Red Blood Cell Phantom Creation*

Red blood cell (RBC) phantoms were created consisting of three layers of agarose, with the middle layer containing 5 % (v/v) red blood cells following previously published methods (Maxwell *et al.*, 2010; Vlaisavljevich *et al.*, 2013a; Edsall *et al.*, 2021a). A 1 % agarose mixture was generated using the agarose preparation above. Fresh porcine blood was obtained from subjects in an unrelated study and added to an anticoagulant solution of Citrate Phosphate Dextrose Anticoagulant (CPD, Sigma Aldrich Corporation, St. Louis, MO, USA), with a CPD-to-blood ratio of 1:9 mL. Whole blood was separated by centrifugation at 3,000 rpm for 10 min. The plasma and white buffy coat were removed, and the RBCs were saved for addition to the phantom. The RBC phantom was created using an initial layer of agarose poured into the tissue phantom holders at 45 °C. The housing was placed in a refrigerator at 4 °C to allow the base agarose layer to cool and solidify. The remaining solution was kept in the vacuum chamber until it had cooled to 38 °C. 9.5 mL of the respective agarose solutions was combined with the RBCs (5 % v/v) by gentle inversion and poured on top of the chilled solidified agarose layer. The liquid RBC-agarose coated the entire surface before the excess solution was poured out leaving behind a thin layer of the RBC-agarose solution. The phantom was replaced in the refrigerator for five minutes. Once the thin RBC-agarose layer was solidified, the remaining agarose solution without RBCs was poured on top of the first two layers to fill the silicone mold and replaced in the refrigerator for one hour to fully solidify producing a thin layer of RBCs suspended in the center of the clear 1 % agarose phantom.

*Red Blood Cell Ablation*

The histotripsy ablation efficiency was examined and compared using agarose tissue phantoms with embedded RBC layers. RBC ablation can be directly visualized as successive pulses lyse the red blood cells and turn the embedded layer translucent (Maxwell *et al.*, 2010; Lin *et al.*, 2014b). Lesion visualized in RBC phantoms have been shown to closely resemble histology of lesions generated in tissue in previous studies (Maxwell *et al.*, 2010; Lin *et al.*, 2014b). The RBC phantom was oriented at the transducer's focus with the embedded RBC layer parallel to the direction of ultrasound propagation. 2,000 histotripsy pulses were applied at $p-= 31$ MPa to a single point in the RBC layer at 1 Hz PRF. High-speed optical imaging using the FLIR camera and high-speed strobe captured the bubble cloud formed from each pulse and an image of the resulting ablated region between pulses. The bubble cloud images were separated from the ablation images and the ablated areas resulting from each pulse were measured using image thresholding by a custom MATLAB script. Finally, to normalize this data, the ablated area of each image for each frame was divided by the respective mean area of the bubble cloud formed in the RBC phantom for each 3 MHz arrival time.

## Results

*Single Bubble Dynamics*

High-speed optical imaging captured the growth and collapse of the bubble cloud at 5 µs intervals from 1 µs before to 109 µs after pulse arrival (**Fig. 3**). Bubbles are seen to grow and collapse nearly symmetrically with little translation. 20 individual bubbles' growth and collapse events were captured for each 3 MHz arrival time resulting in individual radius vs. time curves for -100 ns, 0 ns, and +100 ns (**Fig. 4A-C**). All individual bubble's (n = 20) growths and collapses were averaged at each time point to produce the mean radius vs. time plot in **Figure 4D** showing each bubble as it reached its maximum radius ($R_{max}$) at a time ($t_{max}$) and tracking its size until it reached total collapse ($t_{c1}$, $t_{c2}$). The mean $R_{max}$ for each 3 MHz arrival time was found to be equal to 260.2±16.3 µm, 289.5±19.9 µm, and 241.87±25.2 µm for -100 ns, 0 ns, and +100 ns respectively. The respective $t_{max}$ to reach $R_{max}$ was found to be equal to 24.5±3.6 µs, 27.6±4.3 µs, 24.0±11.2 µs for -100 ns, 0 ns, and +100 ns 3 MHz arrival times. The bubbles all collapsed from these maximum points to the time of first collapse ($t_{c1}$) at 46.5±4.1 µs, 55±4.5 µs, and 42.8±4.8 µs for -100 ns, 0 ns, and +100 ns before showing a rebound in diameter for all and a second collapse at time ($t_{c2}$) of 70.3±7.4 µs, 85.25±10.0 µs, 63.8±8.2 µs for -100 ns, 0 ns, and +100 ns cases. These results show bubble clouds achieving similar $R_{max}$ values to previously published experimental results for 500 kHz ($R_{max}$= 297.4±50.5 µm, $t_{max}$ =28 µs and $t_c$=72 µs) and total collapse (Vlaisavljevich *et al.*, 2015c).

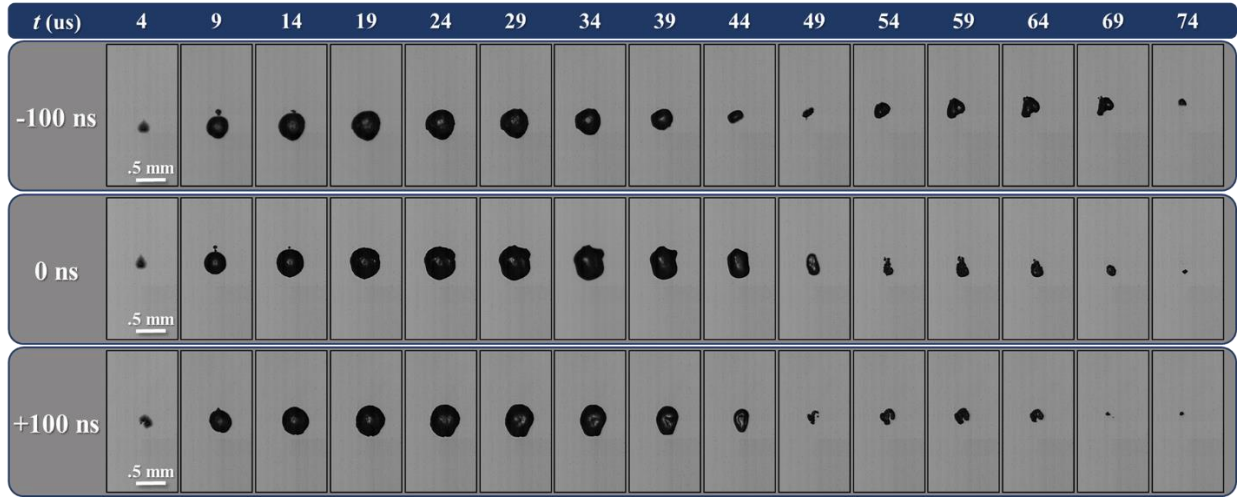

*Figure 3 Optical Single Bubble Size and Dynamics Comparison.* Example optical images shown between 4-74 µs after pulse arrival taken slightly above the respective cavitation thresholds for each 3 MHz arrival time captured by high-speed optical imaging at 5 µs intervals. Ultrasound propagating from bottom to top.

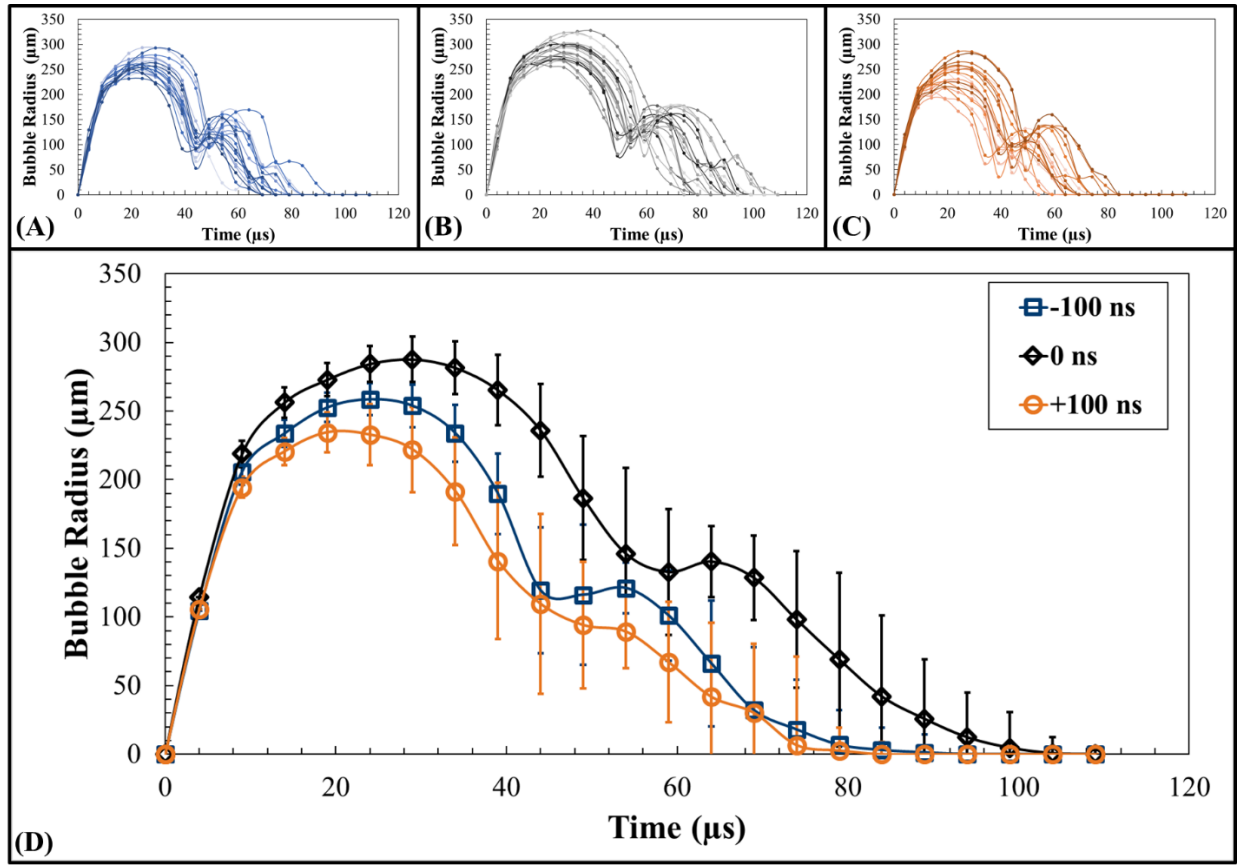

*Figure 4 $R_t$ Curves.* Plots shows all (n = 20) single bubble radius vs time for **(A)** -100 ns, **(B)** 0 ns, and **(C)** +100 ns 3 MHz pulse arrivals as well as **(D)** the mean radius ($R_t$) for single bubble cloud dimensions at 5 µs increments from -1 µs to 109 µs after peak negative pulse arrival. The mean maximum radius ($R_{max}$) was found to be equal to 260.2±16.3µm, 289.5±19.9 µm, and 241.87±25.2 µm at times to maximum radius ($t_{max}$) of 24.5±3.6 µs, 27.6±4.3 µs, 24.0±11.2 µs for -100 ns, 0 ns, and +100 ns cases. The respective times to total collapse ($t_{c2}$) as determined from optical imaging were 70.3±7.4 µs, 85.25±10.0 µs, 63.8±8.2 µs for -100 ns, 0 ns, and +100 ns cases.

*Bubble Cloud Size*

Optical imaging was used to visualize histotripsy bubble clouds generated at *p-* ranging from 22 to 31 MPa at 1 PRF inside agarose tissue phantoms early just after nucleation at 3 µs following *p-* pulse arrival (**Fig. 5**). For all cases, the bubble cloud dimensions were measured in the axial and elevational directions for pressures above the respective cavitation thresholds up to a maximum *p-* of 31 MPa (**Fig. 6**). As hypothesized, cavitation bubbles were observed for all dual-frequency cases when the focal pressure exceeded ~24-26 MPa. For all cases, sharply delineated bubble clouds were observed, with bubble clouds of increasing size containing more bubbles seen with increasing pressure levels beyond the cavitation threshold, matching previous intrinsic threshold histotripsy studies (**Fig. 5**) (Lin *et al.*, 2014b; Maxwell *et al.*, 2013; Vlaisavljevich *et al.*,

2017). The measured bubble cloud areas at *p-* ranging from 26-31 MPa are shown in **Figure 6** with the minimum and maximum mean cloud dimensions and areas for each 3 MHz arrival time summarized in **Table 1**.

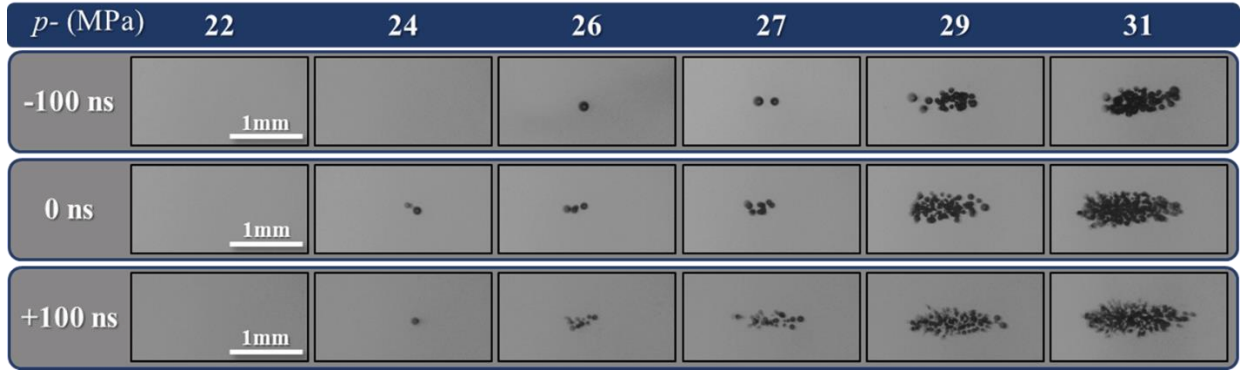

*Figure 5 Bubble Cloud Images.* Optical images captured by the FLIR camera and high-speed strobe of cavitation bubble clouds generated by 1 Hz PRF histotripsy pulses inside 1 % agarose phantoms shown at the same actual scale. Ultrasound propagating left to right.

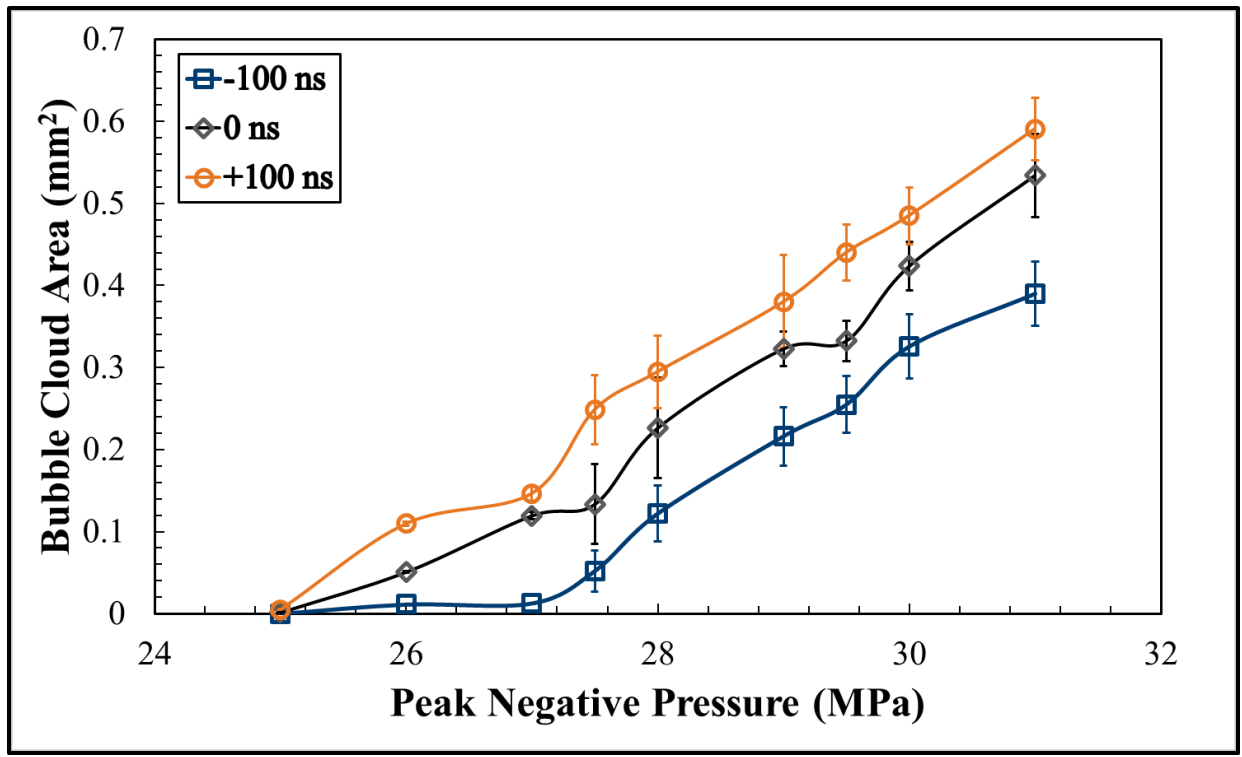

*Figure 6 Bubble Area Cloud Comparison.* Plot shows the mean measured bubble cloud areas calculated by $A = (Axial\ Dimension/2)(Elevational\ Dimension/2)\pi$ from the length and height measurements of each bubble cloud.

*Table 1 Bubble Cloud Dimensional Summary.* Mean bubble cloud lengths, heights and areas plus or minus the respective standard deviations for 26 and 31 MPa for each dual-frequency case.

| 3 MHz Arrival | Axial (mm) | | Elevational (mm) | | Area (mm$^2$) | |
|---|---|---|---|---|---|---|
| | 26 MPa | 31 MPa | 26 MPa | 31 MPa | 26 MPa | 31 MPa |
| -100 ns | 0.14.±0.23 | 1.09±0.09 | 0.10±0.15 | 0.45±0.03 | 0.01±0.03 | 0.39±0.04 |
| 0 ns | 0.34±0.23 | 1.39±0.09 | 0.19±0.12 | 0.49±0.02 | 0.05±0.05 | 0.53±0.04 |
| +100 ns | 0.56±0.16 | 1.55±0.10 | 0.25±0.22 | 0.48±0.02 | 0.11±0.04 | 0.59±0.04 |

Overall, the results indicate that bubble clouds across dual-frequency cases proved intermediate to the previously published dimensions for single-frequency 500 kHz and 3 MHz cases, with the bubble cloud dimensions being closer to previously published dimensions for 3 MHz (Area = 0.01 ± 0.03 mm² at 26 MPa to 0.62 ± 0.03 mm² at 40 MPa) (Edsall *et al.*, 2021a). The 3 MHz arrival time was shown to slightly modulate cloud size with the late arrival (+100 ns) of the 3 MHz producing a larger cloud and the early arrival (-100 ns) producing a smaller cloud than the aligned arrival (0 ns). This difference in cloud size demonstrates some capability for dual-frequency pulsing to further alter cloud size by altering the arrival times of the contributing pulses.

*Bubble Cloud Nucleation and Dynamics*

High-speed optical imaging of the bubble cloud nucleation growth and collapse using the Phantom high-speed camera can be seen in **Figure 7** and is quantified in **Figure 8**. **Figure 7** shows selected clouds generated from a single dual-frequency histotripsy pulse taken at 5 µs intervals from just before peak negative pulse arrival through collapse. Examining these images (**Fig. 7**) along with **Figure 8**, the overall effect of the 3 MHz pulse arrival timing on the bubble clouds dynamics can be assessed. At 4 µs following *p-* arrival, the bubble clou ds have nucleated very similar clouds for all three cases, with only some slight differences in bubble size and bubble cloud area. For example, the 0 ns and +100 ns cases showed similarly sized bubble clouds at this timepoint with mean areas of 0.96±0.04 mm$^2$ and 0.92±0.04 mm$^2$, whereas the -100 ns case produced a slightly smaller cloud of 0.69±0.04 mm$^2$ at this timepoint. Beginning at 9 µs, significant differences in cloud dynamics were observed. Bubble clouds produced by -100 ns and 0 ns continued to grow and covered a greater area that was uniformly aligned over the entire location where the bubbles were initially nucleated. In contrast, the rear portion of the +100 ns cloud (post focal region) had already begun to collapse despite the front portion of the cloud continuing to expand. The total area covered by +100 ns grew to a mean maximum area of 1.03±0.05 mm$^2$ at ~19 µs and began to collapse from this time, while 0 ns and -100 ns continued to grow to 1.47±0.078 mm$^2$ at ~29 µs and 1.2 ±0.12 mm$^2$ at ~34 µs, respectively, before beginning to collapse. These differences were primarily due to the reduced expansion and more rapid collapse of bubbles in the rear portion of the cloud for the later arrival time of the 3 MHz pulse (+100 ns). The collapse time for the entire bubble cloud produced for the 0 ns and -100 ns cases were similar, with the +100 ns cloud showing a slightly earlier collapse time.

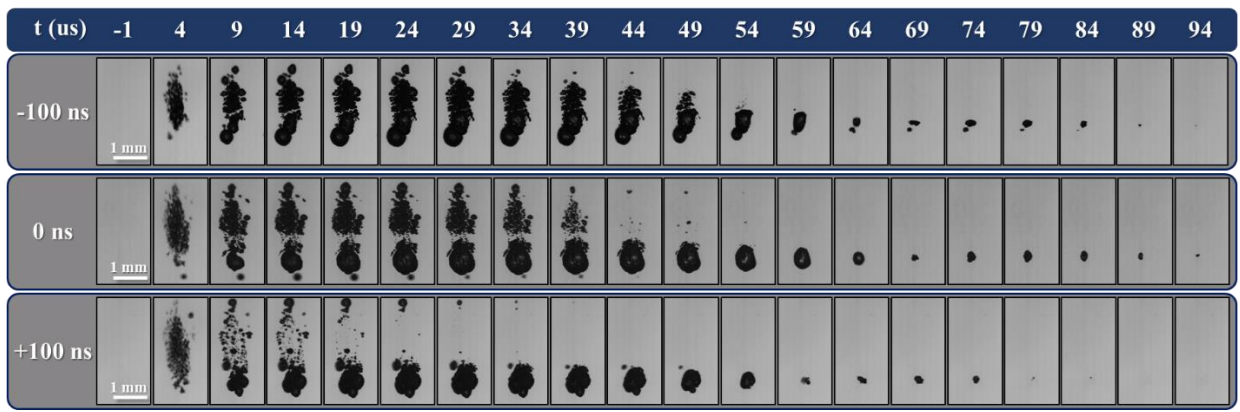

*Figure 7 Bubble Cloud Expansion and Collapse.* Optical images captured by the Phantom high-speed camera of cavitation bubble cloud expansion and collapse inside 1 % agarose phantoms shown at the same actual scale generated by histotripsy pulses applied at 36 MP at 1 Hz PRF. Ultrasound propagating bottom to top.

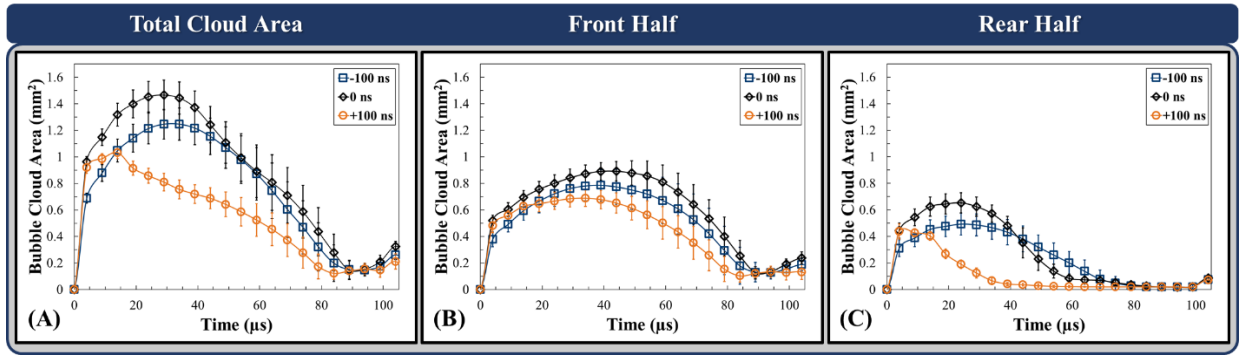

*Figure 8 Bubble Cloud Growth and Collapse Comparison*. Plots show the **(A)** mean measured total area and the mean areas of the **(B)** front and **(C)** rear halves of the bubble cloud at each time point (5 µs intervals) through expansion and collapse measured from binary images of clouds captured by the Phantom high-speed camera at 36 MPa (n = 30).

The observation that the bubble cloud dynamics varied in the front and rear regions of the clouds (**Fig. 7**) led to individual analysis comparing the effects of 3 MHz pulse arrival on the bubble cloud dynamics in the front and rear halves of the cloud (**Fig. 8B/C**). In general, results showed similar behavior in the front regions of the bubble clouds for all cases, with significant changes to the expansion and collapse of bubbles in the rear portion of the clouds (**Fig. 7, Fig 3.8**). The front regions of the bubble clouds were observed to nucleate, grow, and collapse similarly for all cases with each of the arrival times reaching their maximum mean areas of 0.78±0.12 mm$^2$ at ~39 µs, 0.89±0.07 mm$^2$ at ~44 µs, and 0.69±0.06 mm$^2$ at ~34 µs, respectively (**Fig. 8B**). These front regions reached their respective collapse times at ~94 µs, ~89 µs, and ~84 µs for -100 ns, 0 ns, and +100 ns arrival times (**Fig. 8B**). In examining the rear portions of the bubble clouds, results showed significant differences in the bubble growth and collapse dynamics (**Fig. 8C**). For instance, bubble clouds for the -100 ns case achieved a maximum mean rear area of 0.49±0.08 mm$^2$ at a similar time to the front half of the bubble cloud (~39 µs) before collapsing at ~69 µs. In contrast, clouds produced at 0 ns expanded until ~24 µs to reach a maximum mean rear area of 0.65±0.08 mm$^2$ before collapsing at ~59 µs but at a more accelerated rate. This trend was even more apparent in the rapid collapse of rear portion of the bubble cloud for the +100 ns case, which reached a mean maximum area of only 0.44± 0.06 mm$^2$ at ~4 µs before collapse at ~44 µs (**Fig. 8C**). In addition, **Supplemental Figure 1** (**Appendix A**) shows cloud differences in bubble cloud growth between 1-9 µs in greater detail for different clouds captured at 0.5 µs interval by the machine vision FLIR camera and 1 µs strobe. As a note, the number of bubbles in each cloud and the corresponding bubble density analysis could not be effectively collected in this study and compared to our prior single-frequency study due to the extensive bubble overlap resulting from very dense bubble clouds formed with larger bubbles in all dual-frequency cases.

*Ablation in RBC Phantoms*

The results of RBC ablation experiments showed that dual-frequency histotripsy predictably and reproducibly created lesions with clear demarcations between treated and untreated RBC regions (**Fig. 9**). The lesion area increased with increasing pulse number across all cases resulting in well-defined, complete ablations localized to the focal region (**Fig 9, Fig. 10A**). **Figure 10C** shows ablation from the first 100 pulses to better visualize the lesion formation during these initial pulses when the lesion is rapidly forming. The ablation areas were then normalized to the focal area (measured as the mean area of the bubble cloud in the RBC phantom) to quantitatively assess and compare the ablation efficiencies (**Table 2**) (**Fig. 10B/D**). All tested dual-frequency cases produced rapid ablation, achieving >75 % ablation after an average of 37.7±17.5, 69.3±48.4, and 49±11.5 pulses for -100 ns, 0 ns, and +100 ns respectively. The final lesions' mean areas were $A_{-100} = 0.68±0.04$ mm$^2$, $A_0 = 1.09±0.21$ mm$^2$, and $A_{+100} = 1.26±0.11$ mm$^2$ (**Fig. 10A**) reflecting 156±5.7, 148±28.3 %, and 149±13.5 % (**Fig. 10C**) of the focal area after the complete treatment of 2,000 pulses. Closely examining the lesion formation over the first 100 pulses, the lesion areas were $A_{-100} = 0.44±0.01$ mm$^2$, $A_0 = 0.67±0.16$ mm$^2$, $A_{+100} = 0.74±0.004$ mm$^2$ (**Fig. 10B**) reflecting 102±8.1 %, 91.4±22.3 %, and 88±0.49 % (**Fig. 10D**) of the focal area as predicted by the size of the bubble cloud. The rates presented here show an increased ablation efficiency for all dual-frequency cases compared to previously published results for 500 kHz (>250 pulses), 1 MHz (>800 pulses), and 3 MHz (>1000 pulses) single-frequency histotripsy pulsing when removing the respective bubble cloud areas (Edsall *et al.*, 2021a).

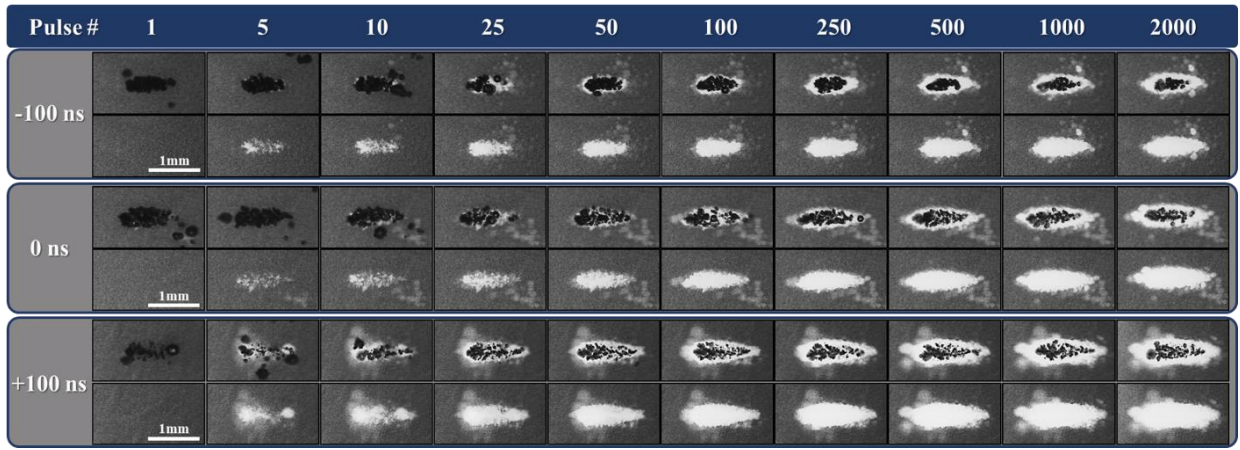

*Figure 9 RBC Ablation.* Images captured by optical imaging using the FLIR camera show the cavitation bubble cloud (dark) and histotripsy lesions (white) generated in RBC phantoms (grey) by histotripsy pulsing applied at 31 MPa at 1 Hz PRF propagating left to right for each 3 MHz arrival time

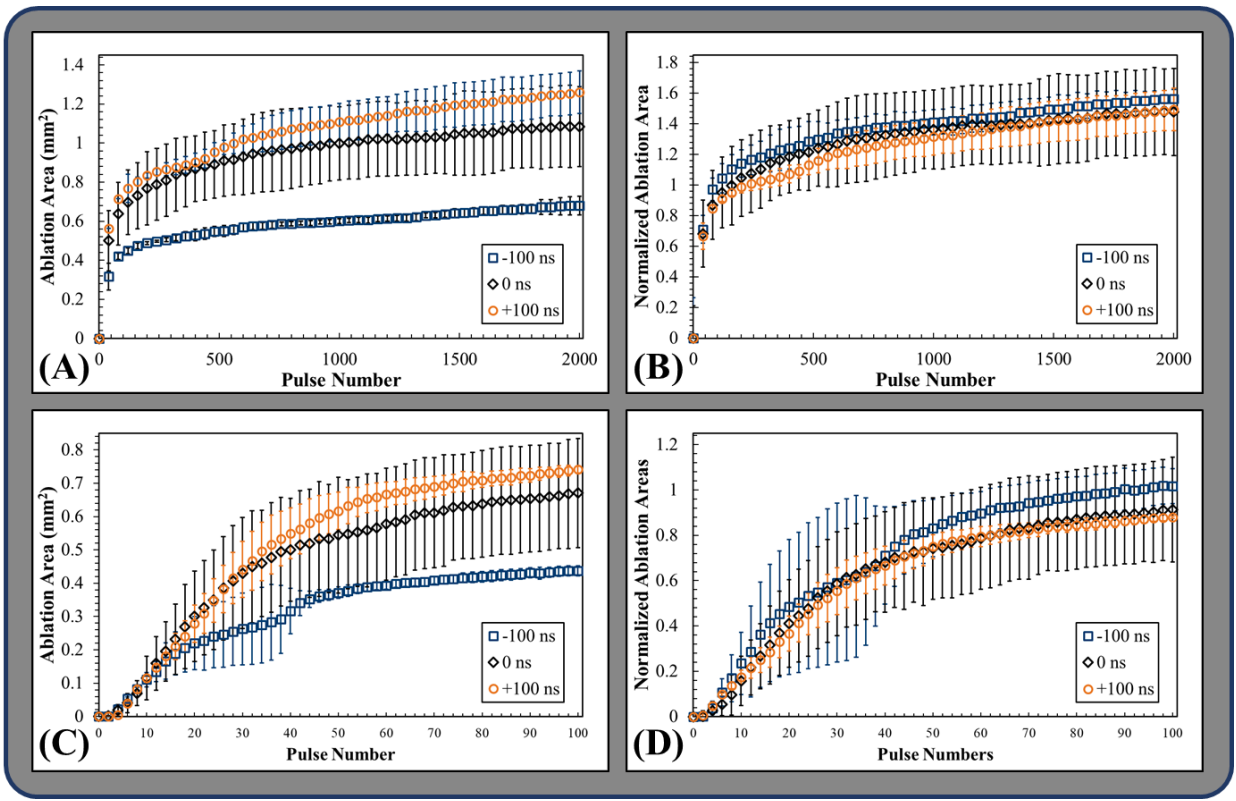

*Figure 10 RBC Ablation Plots.* Plots show the mean and standard deviation of **(A)** the measured ablation area formed after every 40th pulse in the RBC agarose gel phantoms by 1Hz PRF histotripsy pulsing at 31 MPa, **(B)** the measured ablation area for every 40th pulse normalized to the respective mean measured clouds areas formed in the RBCs, and **(C)** the ablation areas and **(D)** normalized areas following every other pulse of the first hundred pulses.

*Table 2 Dual-Frequency Ablation Efficiency Summary.* The table shows the mean number of pulses plus or minus the standard deviation required to achieve the specified percent ablation of the respective normalized focal areas from all dual-frequency histotripsy cases applied at *p-* was 31 MPa.

| Percent of Measured Mean Bubble Cloud Area | 25% | 50% | 75% | 100% |
|---|---|---|---|---|
| 3 MHz Arrival | Number of Pulses | | | |
| -100 ns | 19±16.5 | 24.7±15.1 | 37.7±17.5 | 104.3±48.8 |
| 0 ns | 16.3±7.6 | 29.6±16.8 | 69.3±48.4 | 241.3±174.1 |
| +100 ns | 14±3 | 26.6±4.7 | 49±11.5 | 256.3±114.2 |

# Discussion

This work builds on prior studies that have proposed dual-frequency histotripsy methods for generating more precise ablation of deep targets (Lin *et al.*, 2015; Lin *et al.*, 2014a) and providing precise control over bubble size to design more tissue selective treatment methods (Mancia *et al.*, 2019). In this study, the bubble cloud behavior and ablative efficiency of intrinsic threshold histotripsy generated using dual-frequency pulsing (500 kHz:3 MHz with a 1:1 pressure ratio) were evaluated. As expected, the results from this study showed that the bubble cloud dimensions, bubble size, and bubble density observed for the dual-frequency cases were between those previously reported for the individual 500 kHz and 3 MHz cases (Vlaisavljevich *et al.*, 2015c). Single bubble expansion and collapse results showed that for all dual-frequency cases the single bubbles' radii and collapse times were intermediate to those of previously published results for 3 MHz and 500 kHz single-frequencies but more closely approached the those of 500 kHz (Vlaisavljevich *et al.*, 2015c). These findings suggest that the individual bubbles are more strongly influenced by the low-frequency pulse resulting in larger bubbles with only slightly reduced $R_{max}$ and collapse time results. This also confirms the results of prior work (Lin *et al.*, 2014a) by showing insignificant effects of 3 MHz pulse arrival time on individual bubble size. At higher pressures, the results from this study showed the significant benefits of using dual-frequency histotripsy for generating dense, well-confined bubble clouds that can harness the respective features of low (larger bubble expansion) and high (higher bubble density, more confined bubble cloud) frequency pulsing to achieve more efficient tissue ablation. Optical imaging results of cloud formation show delineated, densely populated bubble clouds of a size intermediate to previously published results for the contributing single frequencies of 500 kHz and 3 MHz, but more closely aligned with the higher-frequency's cloud size (Edsall *et al.*, 2021a; Vlaisavljevich *et al.*, 2015c). These results confirm findings of previous studies showing bubble cloud dimensions predictably increase in size and number of bubbles with increasing pressure (Lin *et al.*, 2014b; Maxwell *et al.*, 2013; Vlaisavljevich *et al.*, 2017) and further support the predictability and reproducibility of using intrinsic threshold histotripsy for precise ablative applications. Although bubble density couldn't be quantified in this study due to the very high density cloud that consisted of rapidly expanding bubbles observed in this dual-frequency case, the qualitative observations of this study highlight how dual-frequency pulsing can be utilized to optimize the trade-offs in bubble density and bubble expansion that were observed in our prior frequency study (Edsall *et al.*, 2021a). These findings suggest that dual-frequency histotripsy can be used to generate more precise ablation without the normal trade-offs of using higher frequency, such as reduced bubble size (Vlaisavljevich *et al.*, 2015c; Lin *et al.*, 2014a), increased tissue attenuation (Vlaisavljevich *et al.*, 2015b), and reduced ablation efficiency (Lin *et al.*, 2014a; Edsall *et al.*, 2021a).

Results from the bubble cloud growth and collapse analysis showed dual-frequency bubble clouds exhibited typical intrinsic threshold nucleation, with well-defined bubble clouds covering the region of the focus above the cavitation threshold. However, an interesting finding of this study showed that the timing of the 3 MHz pulse significantly affected the dynamics of the bubble clouds, particularly the rear portion of the bubble clouds. While the front portions of the bubble clouds showed similar rates of growth and collapse, the rear portions of the bubble cloud showed distinctly different dynamics depending upon the arrival time of the 3 MHz pulse. For instance, the early arrival of the 3 MHz pulses produced a bubble cloud with bubbles expanding to similar sizes with similar collapse times in the front and rear portions of the cloud. In contrast, the aligned (0 ns) and delayed (+100 ns) arrival of the 3 MHz resulted in decreasing growth of the rear half of the bubble cloud relative to the front of the bubble cloud. This change in bubble expansion due to pulse arrival indicates the sensitivity of pulse alignment in determining the ablative capacity of dual-frequency histotripsy. These findings suggested that an earlier arrival time for the higher frequency pulse allows for more uniform ablation of the entire focal area when using dual-frequency pulsing methods. We hypothesize that aligning the 3 MHz pulse with the initial arrival of the longer 500 kHz pulse allows for the bubbles within the cloud to be nucleated at an earlier point of the 500 kHz pulse, allowing for the extended tensile phase of the pulse to grow a higher percentage of the nucleated bubbles to a larger size. This hypothesis is based upon prior experimental (Lin *et al.*, 2014a; Vlaisavljevich *et al.*, 2015c) and theoretical (Mancia *et al.*, 2017) findings that show low frequencies produce greater bubble expansion due to the longer duration of the applied *p-*. In contrast, results from the aligned and late arrival cases for the 3 MHz pulse show that only the bubbles in the front of the cloud appear to be expanded due to the 500 kHz pulse, whereas bubbles in the rear portion of the cloud appear to be experiencing more limited expansion behavior that is typical of the higher frequency 3 MHz pulse.

In the final part of this study, ablation experiments showed that dual-frequency pulsing resulted in precise ablation of the focal areas with even greater efficiency than what was previously observed for any single frequency ablation in our prior study, including the most efficient ablation that was observed for 500 kHz pulsing. The lesions produced in the RBC phantoms in this study were generated in fewer pulses than what was previously reported for 500 kHz or 3 MHz, regardless of 3 MHz arrival time. The arrival time for the high frequency 3 MHz pulse was shown to impart a greater degree of lesion control with an earlier arrival (-100 ns) resulting in a smaller, more confined lesion that was produced at a faster rate than what was observed for the late (+100 ns) arrival time that created a larger lesion at a slower rate. This finding aligns well with the observed differences in bubble cloud nucleation and dynamics described in the previous section. For instance, comparing the lesion formation in **Figure 9** to the data from the bubble cloud dynamics, it can be seen that the late arrival (+100 ns) lesions were formed more rapidly toward the fronts of the clouds while the rears of the lesions were slower to form, likely due to the reduced growth of bubbles toward the back of the cloud (**Fig. 7**). In both the aligned and late arrival (+100 ns) data, ablation clearly favors the regions of the cloud with the greater degree of bubble expansion, supporting prior work demonstrating the significance of bubble expansion in histotripsy ablative capacity (Edsall *et al.*, 2021b; Mancia *et al.*, 2020; Mancia *et al.*, 2019; Vlaisavljevich *et al.*, 2016). Altogether, the ablation experiments show the ability of dual-frequency to rapidly ablate tissue with high precision and efficiency, with the ability to modulate these properties by changing the pulse arrival times of the respective frequencies, thereby impacting the uniformity and efficiency of the lesion formation. To completely unlock the potential benefits of dual-frequency pulsing, future work should explore this concept further by investigating the ability to modulate bubble cloud size, bubble cloud density, and bubble expansion across a much wider range using larger dual-frequency transducers that change more parameters including array geometry, pulse arrival times, and the relative pressure ratio produced by the respective high and low frequency components.

Based on all of the prior studies investigating intrinsic threshold histotripsy, our working hypothesis is that histotripsy the portion of tissues ablated by the formation of the bubble cloud is determined by the dimensions of the bubble cloud while the ablation efficiency is determined from the integrated effects of the size and rate of bubble expansion and the density of the bubbles within the bubble cloud. The results in this study suggest that it may be possible to independently manipulate cloud size, bubble size, and bubble density using dual-frequency pulsing methods by selectively altering the frequency components, the transducer geometry, pressure ratios, and other properties. Although further work is needed to fully test this possibility, the results from this study suggest that the bubble cloud dimensions are influenced more strongly by the higher frequency, which provides a constraint on the region of the focus above the intrinsic threshold. In contrast, bubble size appears to be influenced more strongly by the lower frequency components of the waveform that enable enhanced bubble growth, particularly when the bubbles are nucleated at an early time point with respect to the lower frequency pulse. Future work will more closely examine these possibilities through a comprehensive parametric assessment of transducer geometry, array configuration, pulse timing, and *p-* ratios to provide a more complete understanding of the effects of each parameter on bubble cloud characteristics and the resulting ablative efficiency.

## Conclusion

The results of this study show the potential to modulate bubble cloud and ablation characteristics through dual- or multi- frequency histotripsy pulsing. All dual-frequency pulsing cases produced dense bubble clouds with dimensions intermediate to previously published results for its component frequencies (500 kHz:3 MHz), with the cloud size more closely reflecting the beam profile from the high-frequency component. Single-bubble R-T curves showed the dimensions of single bubbles produced near the threshold more closely reflected the expansion of bubbles produced by the low-frequency component in prior work. Examining the changes in bubble cloud behavior showed the arrival of the high frequency pulse significantly altered the growth and collapse dynamics for the rear portions of the cloud, suggesting that an earlier arrival time for the higher frequency pulse should be used to produce uniform and efficient histotripsy ablation using these methods. This hypothesis was confirmed in a final set of RBC ablation experiments that showed all dual-frequency cases were capable of creating complete precise lesions, with the highest ablation efficiency observed when the higher frequency 3 MHz pulse arrived at an early time point with respect to the lower frequency 500 kHz pulse. These findings support prior research showing the stress and strain from maximum bubble radius is fundamental to ablation efficiency and suggests the potential for altering the rate and extent of ablation through the application of dual-frequency pulse modulation. Overall, this study demonstrates the potential of dual-frequency to increase control over histotripsy bubble cloud characteristics, behavior, and ablation efficiency and thereby potentially improve the efficiency, safety, and efficacy of histotripsy-based therapies.

## Acknowledgments

This work was funded in part by grants from the NIH (R01CA262474, R01CA269811-01). The authors would like to thank the Virginia Tech Institute for Critical Technology and Applied Science (ICTAS) and the Department of Biomedical Engineering and Mechanics for their support of this work. We would also like to thank Dr. Coutier-Delgosha, Roshan Kumar Subramanian, and the Virginia Tech Department of Mechanical Engineering for the use of the Phantom high-speed camera used in this study. Drs. Eli Vlaisavljevich and Timothy L. Hall have financial relationships and research partnerships with HistoSonics, LLC.

All of the authors contributed significantly to the development and execution of this study. EV and CE conceived the initial ideas for this work. CE and TH contributed to the focused ultrasound transducer design and construction and the development of the experimental setups. CE and LH collected and analyzed all cavitation and red blood cell lesion data captured by optical imaging. CE also collected, conceived, developed, and performed methods for analyzing single-bubble and bubble cloud growth and collapse data collected by highspeed imaging. CE led the writing of this manuscript. All authors contributed critically to the interpretation of results, drafting of the manuscript and gave final approval for this submission.

# References

Bader K B, Vlaisavljevich E and Maxwell A D 2019 For Whom the Bubble Grows: Physical Principles of Bubble Nucleation and Dynamics in Histotripsy Ultrasound Therapy *Ultrasound Med Biol* **45** 1056-80

Duck F A 1990 *Physical properties of tissue: a comprehensive reference book.* : Academic Press

Edsall C, Ham E, Holmes H, Hall T L and Vlaisavljevich E 2021a Effects of frequency on bubble-cloud behavior and ablation efficiency in intrinsic threshold histotripsy *Physics in Medicine & Biology* **66** 225009

Edsall C, Khan Z M, Mancia L, Hall S, Mustafa W, Johnsen E, Klibanov A L, Durmaz Y Y and Vlaisavljevich E 2021b Bubble cloud behavior and ablation capacity for histotripsy generated from intrinsic or artificial cavitation nuclei *Ultrasound Med Biol* **47** 620-39

Hall T L, Hempel C R, Wojno K, Xu Z, Cain C A and Roberts W W 2009 Histotripsy of the prostate: dose effects in a chronic canine model *Urology.* **74** 932-7.

Lin K-W, Hall T L, Xu Z and Cain C A 2015 Histotripsy lesion formation using an ultrasound imaging probe enabled by a low-frequency pump transducer *Ultrasound Med Biol* **41** 2148-60

Lin K W, Duryea A P, Kim Y, Hall T L, Xu Z and Cain C A 2014a Dual-beam histotripsy: a low-frequency pump enabling a high-frequency probe for precise lesion formation *IEEE Trans Ultrason Ferroelectr Freq Control* **61** 325-40

Lin K W, Kim Y, Maxwell A D, Wang T Y, Hall T L, Xu Z, Fowlkes J B and Cain C A 2014b Histotripsy beyond the intrinsic cavitation threshold using very short ultrasound pulses: microtripsy *IEEE Trans Ultrason Ferroelectr Freq Control* **61** 251-65

Mancia L, Rodriguez M, Sukovich J, Xu Z and Johnsen E 2020 Single–bubble dynamics in histotripsy and high–amplitude ultrasound: Modeling and validation *Physics in Medicine & Biology* **65** 225014

Mancia L, Vlaisavljevich E, Xu Z and Johnsen E 2017 Predicting Tissue Susceptibility to Mechanical Cavitation Damage in Therapeutic Ultrasound *Ultrasound Med Biol* **43** 1421-40

Mancia L, Vlaisavljevich E, Yousefi N, Rodriguez M, Ziemlewicz T J, Lee F T, Henann D, Franck C, Xu Z and Johnsen E 2019 Modeling tissue-selective cavitation damage *Physics in Medicine & Biology* **64** 225001

Maxwell A D, Cain C A, Hall T L, Fowlkes J B and Xu Z 2013 Probability of cavitation for single ultrasound pulses applied to tissues and tissue-mimicking materials *Ultrasound Med Biol* **39** 449-65

Maxwell A D, Wang T Y, Yuan L, Duryea A P, Xu Z and Cain C A 2010 A tissue phantom for visualization and measurement of ultrasound-induced cavitation damage *Ultrasound Med Biol* **36** 2132-43

Parsons J E, Cain C A, Abrams G D and Fowlkes J B 2006 Pulsed cavitational ultrasound therapy for controlled tissue homogenization *Ultrasound in Medicine and Biology* **32** 115-29

Roberts W W, Hall T L, Ives K, Wolf J S, Jr., Fowlkes J B and Cain C A 2006 Pulsed cavitational ultrasound: a noninvasive technology for controlled tissue ablation (histotripsy) in the rabbit kidney *J Urol* **175** 734-8

Roberts W W, Teofilovic D, Jahnke R C, Patri J, Risdahl J M and Bertolina J A 2014 Histotripsy of the prostate using a commercial system in a canine model *J Urol* **191** 860-5

Vidal-Jove J, Serres X, Vlaisavljevich E, Cannata J, Duryea A, Miller R, Merino X, Velat M, Kam Y and Bolduan R 2022 First-in-man histotripsy of hepatic tumors: the THERESA trial, a feasibility study *International Journal of Hyperthermia* **39** 1115-23

Vlaisavljevich E, Aydin O, Durmaz Y Y, Lin K W, Fowlkes B, ElSayed M and Xu Z 2015a Effects of Ultrasound Frequency on Nanodroplet-Mediated Histotripsy *Ultrasound Med Biol*

Vlaisavljevich E, Durmaz Y Y, Maxwell A, Elsayed M and Xu Z 2013a Nanodroplet-mediated histotripsy for image-guided targeted ultrasound cell ablation *Theranostics* **3** 851-64

Vlaisavljevich E, Gerhardson T, Hall T and Xu Z 2017 Effects of f-number on the histotripsy intrinsic threshold and cavitation bubble cloud behavior *Phys Med Biol* **62** 1269-90

Vlaisavljevich E, Kim Y, Allen S, Owens G, Pelletier S, Cain C, Ives K and Xu Z 2013b Image-Guided Non-invasive Ultrasound Liver Ablation Using Histotripsy: Feasibility Study in an In Vivo Porcine Model *Ultrasound Med Biol*

Vlaisavljevich E, Lin K W, Maxwell A, Warnez M T, Mancia L, Singh R, Putnam A J, Fowlkes B, Johnsen E, Cain C and Xu Z 2015b Effects of ultrasound frequency and tissue stiffness on the histotripsy intrinsic threshold for cavitation *Ultrasound Med Biol* **41** 1651-67

Vlaisavljevich E, Lin K W, Warnez M, Singh R, Mancia L, Putnam A, Johnsen E, Cain C and Xu Z 2015c Effects of Tissue Stiffness, Ultrasound Frequency, and Pressure on Histotripsy-induced Cavitation Bubble Behavior *Phys Med Biol*

Vlaisavljevich E, Maxwell A, Mancia L, Johnsen E, Cain C and Xu Z 2016 Visualizing the Histotripsy Process: Bubble Cloud-Cancer Cell Interactions in a Tissue-Mimicking Environment *Ultrasound Med Biol*

Vlaisavljevich E, Maxwell A, Warnez M, Johnsen E, Cain C A and Xu Z 2014 Histotripsy-induced cavitation cloud initiation thresholds in tissues of different mechanical properties *IEEE Trans Ultrason Ferroelectr Freq Control* **61** 341-52

Xu Z, Hall T L, Vlaisavljevich E and Lee Jr F T 2021 Histotripsy: the first noninvasive, non-ionizing, non-thermal ablation technique based on ultrasound *International Journal of Hyperthermia* **38** 561-75

Xu Z, Ludomirsky A, Eun L Y, Hall T L, Tran B C, Fowlkes J B and Cain C A 2004 Controlled ultrasound tissue erosion *IEEE Trans Ultrason Ferroelectr Freq Control* **51** 726-36

Yamada H 1973 *Strength of biologic materials.* (New York: Robert E. Kreiger)

Zhang X, Owens G E, Gurm H S, Ding Y, Cain C A and Xu Z 2015 Noninvasive thrombolysis using histotripsy beyond the intrinsic threshold (microtripsy) *IEEE Trans Ultrason Ferroelectr Freq Control* **62** 1342-55